\documentclass[aps,pra,preprint,showpacs,showkeys]{revtex4}
\usepackage{amssymb,bm}
\usepackage{graphicx}
\usepackage{amsmath}
\usepackage{epstopdf}
\usepackage{float}
\allowdisplaybreaks

\begin{document}
\title{High-energy $\mu^+\mu^-$ electroproduction}

\author{P. A. Krachkov}\email{peter_phys@mail.ru}
\author{A. I. Milstein}\email{A.I.Milstein@inp.nsk.su}
\affiliation{Budker Institute of Nuclear Physics, 630090 Novosibirsk, Russia}

\date{\today}

\begin{abstract}
	The cross sections of high-energy $\mu^+\mu^-$ pair production and  paradimuonium production by a relativistic electron in the atomic field are discussed.  The calculation is performed exactly in the parameters of the atomic field. Though the Coulomb corrections to the cross sections, related to the multiple Coulomb  exchange of $\mu^+,\,\mu^-$  and an atom, are negligible,  the Coulomb corrections to the differential cross sections, related to  the electron interaction with an atom, are large. However, the latter  Coulomb corrections to the cross sections integrated over the final electron momentum  are small. Apparently, this effect can easily be observed experimentally. Furthermore, it is shown that the asymmetry of the cross section with respect to the permutation   of the momenta of $\mu^-$ and $\mu^+$ is  large.
\end{abstract}



\maketitle
	
\section{Introduction}

A process of $\mu^+\mu^-$ pair production by a high-energy electron in the atomic field is one of the most important QED processes. Because of its importance, this process has been investigated in many papers \cite{Bhabha2,Racah37,BMV13} in the leading  in the parameter $\eta=Z\alpha$ approximation (in the Born approximation),  where $Z$ is the atomic charge number and $\alpha$ is the fine-structure constant. 
A pair of  $\mu^+$ and $\mu^-$ may be in an unbound state as well as in the bound state (dimuonium).
Dimuonium is now widely discussed \cite{BS61,BL09,BS12,CZ12,EB15,Lamm16,Olsen85,MSS71,HO87,ACS00,GJKKSS98,KKSS99,BNNT} because it is one of the simplest hydrogen-like atoms. This atom is very convenient for testing fundamental laws. Several proposals are currently under consideration. In proposal of Budker Institute~\cite{BDLMS17}, the scheme of a new experiment for the production of dimuonium  in $e^+e^-$ annihilation is suggested. In Jefferson Laboratory, it is planned to produce dimuonium in collision of an electron with a tungsten target  \cite{Hansson14}. In Fermilab, it is planned \cite{REDTOP} to observe  dimuonium in the decay $\eta\rightarrow\gamma (\mu^+\mu^-)$, where $\eta$-meson is produced in collisions of protons with  a beryllium target. An experiment on the production of dimuonium  using the low-energy muon beams is also under consideration \cite{ISS15}.

The Coulomb corrections (the difference between the exact in $\eta$ result and the Born result) to the cross section of the process under consideration are originated  from the interaction of created $\mu^+\mu^-$ pair with the atomic field and from the interaction of an electron with the atomic field. The Coulomb corrections, related to the interaction of created $\mu^+\mu^-$ pair with the atomic field, are strongly suppressed by the atomic form factor \cite{HKS07}, as well as in the case of $\mu^+\mu^-$ photoproduction \cite{IM98}. The Coulomb corrections, related to the interaction of an electron with the atomic field, have not been discussed yet. However, the account for this contribution may significantly modify the differential cross sections of the process. We have found this effect in our recent investigation of $e^+e^-$ electroproduction by a heavy charged particle \cite{KM2017} and  by an ultrarelativistic electron \cite{KM2016,KM2017Arxiv} in the atomic field. In both cases, the Coulomb corrections are significant and  reveal the interesting properties. It has been shown in Ref.~\cite{KM2017}  that the cross section, differential over the angles of a heavy outgoing particle, changes significantly due to the exact account for the interaction of a heavy particle with the atomic field.  However, the cross section integrated over these angles is not affected by this interaction. The same statement is also valid for the process of $e^+e^-$ electroproduction by an ultrarelativistic electron \cite{KM2017Arxiv}. 

In the present paper we investigate the impact of the electron interaction with the atomic field on the cross section  of high-energy $\mu^+\mu^-$ electroproduction. We discuss the electroproduction of unbound  $\mu^+\mu^-$ pair in Sec.~\ref{free} and electroproduction of paradimuonium in  Sec.~\ref{PDM}. It is shown that in both cases the account for the interaction of an electron with the atomic field results in the large Coulomb corrections to the cross section differential over the electron transverse momentum. However, the cross section integrated over these momentum coincides with the Born result.

\section{Electroproduction of unbound $\mu^+\mu^-$ pair.}\label{free}

\begin{figure}[H]
	\centering
	\includegraphics[width=0.4\linewidth]{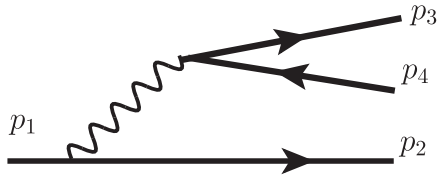}
	\caption{Diagram for the amplitude $T$  of the process $e^-Z\to e^- \mu^+\mu^-Z$. Wavy line denotes the photon propagator, straight lines denote the wave functions in the atomic field.}
	\label{fig:diagrams}
\end{figure}

The differential cross section of high-energy $\mu^+\mu^-$ electroproduction by an electron  in the atomic  field   reads 
\begin{equation}\label{eq:cs}
d\sigma=\frac{\alpha^2}{(2\pi)^8}\,d\varepsilon_3d\varepsilon_4\,d\bm p_{2\perp}\,d\bm p_{3\perp}d\bm p_{4\perp}\,\frac{1}{2}\sum_{\mu_i}|T|^{2}\,,
\end{equation}
where $\bm p_1$ and $\bm p_2$ are initial and final electron momenta, $\bm p_3$ and $\bm p_4$ are momenta of  $\mu^-$ and $\mu^+$, $\varepsilon_{1}=\varepsilon_{2}+\omega$ is the  energy of the  incoming electron, $\omega=\varepsilon_{3}+\varepsilon_{4}$, $\varepsilon_{1,2}=\sqrt{{p}_{1,2}^2+m_e^2}$, $\varepsilon_{3,4}=\sqrt{{p}_{3,4}^2+m_\mu^2}$, $m_e$ is the electron mass, $m_\mu$ is the muon mass, and $\alpha$ is the fine-structure constant,  $\hbar=c=1$.  In Eq.~\eqref{eq:cs} the notation  $\bm X_\perp=\bm X-(\bm X\cdot\bm p_1)\bm p_1/p_1^2$ for any vector $\bm X$ is used, $\mu_i=\pm1$ corresponds to the helicity of the particle with the momentum $\bm p_i$, $\bar{\mu}_i=-\mu_i$. Below  we assume that $\varepsilon_{2}\gg m_e$ and $\varepsilon_{1,3,4}\gg m_\mu$.

To calculate the amplitude $T$ (see Fig.~\ref{fig:diagrams}), we use the quasiclassical approximation \cite{KLM2016} developed in our recent paper \cite{KM2016} for the problem of $e^+e^-$ pair production by a relativistic  electron in the atomic field.
This  approximation is based on the smallness of  angles between the momenta of the final particles and the momentum of the initial particle at high energies.  In this case typical angular momenta, which provide the main contribution to the cross section, are large, $l \sim E/\Delta\gg1$, where $E$ is energy of initial particle and $\Delta$ is the momentum transfer. As a result, the quasiclassical approximation, based on the account for the large angular momentum contribution, becomes applicable.

Following Ref.~\cite{KM2016}, we write the amplitude  $T$ \eqref{eq:cs} in the form
\begin{align}\label{M1new}
&T=T_\perp+T_\parallel\,,\nonumber\\
&T_\perp=-4\pi\sum_{\lambda=\pm}\int \frac{d\bm k\,j_\lambda\,J_\lambda}{(2\pi)^3(\omega^2-k^2+i0)}\,,\nonumber\\
&T_\parallel=-\frac{4\pi}{\omega^2}\int \frac{d\bm k}{(2\pi)^3}j_\parallel\,J_\parallel\,,\nonumber\\
&j_\lambda=\bm j\cdot\bm s_\lambda^*\,,\quad J_\lambda=\bm J\cdot\bm s_\lambda\,,\quad j_\parallel=\bm j\cdot\bm\nu\,,\quad J_\parallel=\bm J\cdot\bm\nu\,.
\end{align}
where  $\bm\nu=\bm k/k$, $\bm s_{\lambda}=(\bm e_x+i\lambda\bm e_y)/\sqrt{2}$, $\bm e_x$ and  $\bm e_y$ are two orthogonal unit vectors perpendicular to $\bm\nu$.
The functions  $\bm j$ and $\bm J$ correspond to  the matrix elements of virtual photon bremsstrahlung and pair production by virtual photon, respectively. We perform the  calculation of these functions  in the same way as it has been done in Refs.~\cite{LMSS2005, KM2015} for  real bremsstrahlung  and in Ref.~\cite{KLM2014} for  pair production by a real photon. 
We obtain 
\begin{align}\label{jlambda}
&j_\lambda=-\frac{A(\bm\Delta)}{\varepsilon_1\varepsilon_2}\Bigg[\delta_{\mu_1\mu_2}(\varepsilon_1\delta_{\lambda\mu_1}+\varepsilon_2\delta_{\lambda\bar\mu_1})
\left(\bm s_\lambda^*, \frac{\bm p_2}{ D_1}+ \frac{\bm p_1}{ D_2}\right)\,\nonumber\\
&+\frac{1}{\sqrt{2}}\delta_{\mu_1\bar\mu_2}\delta_{\lambda\mu_1}m_e\omega\mu_1\left(\frac{1}{D_1}+\frac{1}{D_2}\right)\Bigg]\,,\nonumber\\
&j_\parallel=-A(\bm\Delta)\delta_{\mu_1\mu_2}\left(\frac{1}{D_1}+\frac{1}{D_2}\right)\,,\nonumber\\
&A(\bm\Delta)=-\frac{i}{\Delta_{\perp}^2}\int d\bm r\,\exp[-i\bm\Delta\cdot\bm r-i\chi(\rho)]\bm\Delta_{\perp}\cdot\bm\nabla_\perp V(r)\,,\nonumber\\
&\chi(\rho)=\int_{-\infty}^\infty dz\,V(\sqrt{z^2+\rho^2})\,,\nonumber\\
&D_1=\frac{\Delta_{\perp}^2}{2\varepsilon_1}+\bm n_1\cdot\bm\Delta-i0\,,\quad D_2=\frac{\Delta_\perp^2}{2\varepsilon_2}-\bm n_2\cdot\bm\Delta-i0\,, \nonumber\\
&\bm\Delta=\bm k+\bm p_2-\bm p_1\,,\quad  \bm n_i=\bm p_i/p_i\,,
\end{align}
where $V(r)$ is the atomic potential.
At $R^{-1}\gg\Delta_\perp\gg\max(\Delta_\parallel,r_{scr}^{-1})$, where $\Delta_\parallel=\bm\Delta\cdot\bm\nu$,  $r_{scr}$ is a screening radius, and $R$ is the radius of a nucleus, the function $A(\bm \Delta)$ is independent of the potential shape \cite{LMSS2005} and has the following form
\begin{align}\label{ACJCas}
&A_{as}(\bm\Delta)=-\frac{4\pi\eta (L\Delta_\perp)^{2i\eta}\Gamma(1-i\eta)}{\Delta_\perp^2\Gamma(1+i\eta)}\,,
\end{align}
where  $\Gamma(x)$ is the Euler $\Gamma$ function, a specific value of $L\sim \max(\Delta_\parallel,r_{scr}^{-1})$ is irrelevant because the factor $L^{2i\eta}$ disappears in $|T|^2$.
At $\Delta_\perp\lesssim\max(\Delta_\parallel,r_{scr}^{-1})$,  the function $A(\bm \Delta)$ strongly depends on  $\Delta_\parallel$ and on the shape of the atomic potential  \cite{LMSS2005}.

Note that the main contribution to the Coulomb corrections to the cross section of $\mu^+\mu^-$ photoproduction is given by the impact parameter $\rho\sim \lambda_\mu\ll R$, where $\lambda_\mu=1/m_\mu$. Thus, the Coulomb corrections to the cross section are strongly suppressed by the form factor, and below we use the Born result for the matrix element $J_\lambda$ and $J_\parallel$  of $\mu^+\mu^-$ photoproduction by a virtual photon:

\begin{align}\label{Jlambda}
&J_\lambda=J_\lambda^{(0)}+J_\lambda^{(1)}\,,\quad  J_\parallel=J_\parallel^{(0)}+J_\parallel^{(1)}\,, \nonumber\\
&J_\lambda^{(0)}=\dfrac{(2\pi)^3}{\varepsilon_3\varepsilon_4}\delta(\bm p_3+\bm p_4-\bm k)\left[\delta_{\mu_3\bar\mu_4}\left(\bm s_\lambda, \delta_{\lambda\mu_3}\varepsilon_3\bm p_4+\delta_{\lambda\mu_4}\varepsilon_4\bm p_3\right)
-\frac{1}{\sqrt{2}}\delta_{\mu_3\mu_4}\delta_{\lambda\mu_3}m_\mu\omega\mu_3\right]\,,\nonumber\\
&J_\lambda^{(1)}=-\frac{8\pi\eta F(\Delta^2)}{\omega\Delta^2m_\mu^2}
\left\{\delta_{\mu_3\bar\mu_4} \big[\varepsilon_3\,\delta_{\mu_3\lambda}\bm s_{\mu_3} -\varepsilon_4 \delta_{\mu_3\lambda}\bm s_{\mu_4}\big]\cdot(\xi_{1}\bm p_{3}+\xi_{2}\bm p_{4}) +\delta_{\mu_3\mu_4}\delta_{\mu_3\lambda}\frac{m_e\omega\mu_3}{\sqrt{2}}(\xi_{1}-\xi_{2})\right\}\,,\nonumber\\
&J_\parallel^{(0)}=(2\pi)^3\delta(\bm p_3+\bm p_4-\bm k)\delta_{\mu_3\bar\mu_4}\,,\quad
J_\parallel^{(1)}=\frac{8\pi\eta\varepsilon_3\varepsilon_4 F(\Delta^2)}{\omega^2\Delta^2 m_\mu^2}
 (\xi_{1}-\xi_{2})\delta_{\mu_3\bar\mu_4}\,,\nonumber\\
&\xi_{1}=\frac{m_\mu^2}{m^2_\mu+\varepsilon_3\varepsilon_4(\omega^2-k^2)/\omega^2+\varepsilon_3^2 \theta_{3k}^2}\,,\quad \xi_{2}=\frac{m_\mu^2}{m^2_\mu+\varepsilon_3\varepsilon_4(\omega^2-k^2)/\omega^2+\varepsilon_4^2 \theta_{4k}^2}\,,\nonumber\\
& \bm\Delta= \bm p_3 + \bm p_4 - \bm  k\,,
\end{align}
where $F(\Delta^2)$ is the atomic form factor, $\theta_{3k}$ and $\theta_{4k}$ are the angles between the momenta $\mu^-$, $\mu^+$ and the virtual photon momentum $\bm k$.

Then we substitute Eqs.~\eqref{jlambda} and \eqref{Jlambda} in Eq.~\eqref{M1new}, take the integral   over $\bm k$,  and write the amplitudes $T_\perp$ and $T_\parallel$ in the form
\begin{align}\label{T01}
&T_\perp=T_\perp^{(0)}+T_\perp^{(1)}\,,\quad T_\parallel=T_\parallel^{(0)}+T_\parallel^{(1)}\,.
\end{align}
The terms $T_\perp^{(0)}$ and   $T_\parallel^{(0)}$ read
\begin{align}\label{T0}
&T_\perp^{(0)}=\frac{8\pi A(\bm\Delta_0)\delta_{\mu_1\mu_2}}{m_\mu^2+\zeta^2}\Big\{\delta_{\mu_3\bar\mu_4}\Big[\frac{\varepsilon_3}{\omega}(\bm s_{\mu_3}^*\cdot \bm X)(\bm s_{\mu_3}\cdot\bm\zeta)(\varepsilon_1\delta_{\mu_1\mu_3}+\varepsilon_2\delta_{\mu_1\mu_4})\nonumber\\
&-\frac{\varepsilon_4}{\omega}(\bm s_{\mu_4}^*\cdot \bm X)(\bm s_{\mu_4}\cdot\bm\zeta) (\varepsilon_1\delta_{\mu_1\mu_4}+\varepsilon_2\delta_{\mu_1\mu_3})\Big]+\frac{m_\mu\mu_3}{\sqrt{2}}\delta_{\mu_3\mu_4}(\bm s_{\mu_3}^*\cdot\bm X)(\varepsilon_1\delta_{\mu_3\mu_1}+\varepsilon_2\delta_{\mu_3\bar\mu_1})\Big\}\,,\nonumber\\
&T_\parallel^{(0)}=16\pi A(\bm\Delta_0)\delta_{\mu_1\mu_2}\delta_{\mu_3\bar\mu_4}\frac{\bm\theta_{21}\cdot\bm\Delta_{0\perp}}{D^2 }\,,
\end{align}
where
\begin{align}\label{T0not}
&\bm X=\frac{\bm\Delta_{0\perp}}{\varepsilon_1\varepsilon_2 D}-\frac{2(\bm \theta_{21}\cdot\bm\Delta_{0\perp})\bm\theta_{21}}{D^2}\,,\nonumber\\
& D=\frac{\omega^2}{\varepsilon_3\varepsilon_4}(m_\mu^2+\zeta^2)+\frac{m_e^2\omega^2}{\varepsilon_1\varepsilon_2}+\frac{\varepsilon_1}{\varepsilon_2}p_{2\perp}^2
\,,\quad\bm\zeta=\frac{\varepsilon_3\varepsilon_4}{\omega}\bm\theta_{34}\,,\nonumber\\
& \bm\Delta_0=\bm p_2+\bm p_3+\bm p_4-\bm p_1\,,\quad \bm{\theta}_{ij}=\frac{\bm p_{i\perp}}{\varepsilon_i}-
\frac{\bm p_{j\perp}}{\varepsilon_j}\,,\nonumber\\
&\Delta_{0\parallel}=-\frac{1}{2}\left[\frac{m_\mu^2\omega}{\varepsilon_3\varepsilon_4}+\frac{m_e^2\omega}{\varepsilon_1\varepsilon_2}+\varepsilon_2\theta_{21}^2+\varepsilon_3\theta_{31}^2+\varepsilon_4\theta_{41}^2\right]\,.\end{align}
The terms $T_\perp^{(0)}$ and   $T_\parallel^{(0)}$ correspond to the amplitudes of  electroproduction of $\mu^+\mu^-$ pair non-interacting with the atomic field.

We perform the calculation of $T_\perp^{(1)}$ and $T_\parallel^{(1)}$ as in Ref.~\cite{KM2016}. Then we have 
\begin{align}\label{T1C}
&T_\perp^{(1)}=\frac{8i\eta\varepsilon_1}{\omega} \int\frac{d\bm\Delta_\perp\, A(\bm\Delta_\perp) F(Q^2)}{Q^2 \,(m_e^2\omega^2+\varepsilon_1^2Y^2)}
{\cal M}\,, \nonumber\\
&{\cal M}=-\frac{\delta_{\mu_1\mu_2}\delta_{\mu_3\bar\mu_4}}{\omega} \big[ \varepsilon_1(\varepsilon_3 \delta_{\mu_1\mu_3}-\varepsilon_4 \delta_{\mu_1\mu_4})
(\bm s_{\mu_1}^*\cdot\bm Y)(\bm s_{\mu_1}\cdot\bm I_1)\,\nonumber\\
&+\varepsilon_2(\varepsilon_3 \delta_{\mu_1\bar\mu_3}-\varepsilon_4 \delta_{\mu_1\bar\mu_4})(\bm s_{\mu_1}\cdot\bm Y)(\bm s_{\mu_1}^*\cdot\bm I_1)  \big]+\delta_{\mu_1\bar\mu_2}\delta_{\mu_3\bar\mu_4}\frac{m_e\omega\mu_1}{\sqrt{2}\varepsilon_1 }(\varepsilon_3 \delta_{\mu_1\mu_3}-\varepsilon_4 \delta_{\mu_1\mu_4})(\bm s_{\mu_1}
\cdot\bm I_1)\nonumber\\
&+\delta_{\mu_1\mu_2}\delta_{\mu_3\mu_4}\frac{m_\mu\mu_3}{\sqrt{2}}(\varepsilon_1 \delta_{\mu_1\mu_3}+\varepsilon_2 \delta_{\mu_1\bar\mu_3})(\bm s_{\mu_3}^*\cdot\bm Y)I_0 
-\frac{m_em_\mu\omega^2}{2\varepsilon_1}\delta_{\mu_1\bar\mu_2}
\delta_{\mu_3\mu_4}\delta_{\mu_1\mu_3}I_0\,,\nonumber\\
&T_\parallel^{(1)}=-\frac{8i\eta\varepsilon_3\varepsilon_4}{\omega^3} \int \frac{d\bm\Delta_\perp\, A(\bm\Delta_\perp) F(Q^2)}{Q^2}\,I_0
\delta_{\mu_1\mu_2}\delta_{\mu_3\bar\mu_4}\,,
\end{align}
Here
\begin{align}\label{eq:Aperp}
&A(\bm\Delta_\perp)=i\int d\bm\rho\,\exp[-i\bm\Delta_\perp\cdot\bm\rho-i\chi(\rho)]\,,
\end{align}
is the function $A(\bm\Delta)$, see Eq.~\eqref{jlambda}, at $\Delta_\parallel=0$. The integration over $\bm\Delta_\perp$ in Eqs.~\eqref{T1C} is performed over two-dimensional vectors perpendicular to $z$-axis. The following notations are used in Eqs.~\eqref{T1C}
\begin{align}\label{T1Cnot}
&M^2=m_\mu^2+\frac{\varepsilon_3\varepsilon_4}{\varepsilon_1\varepsilon_2}m_e^2
+\frac{\varepsilon_1\varepsilon_3\varepsilon_4}{\varepsilon_2\omega^2}Y^2\,,\quad \bm Y=\bm\Delta_\perp-\varepsilon_2\bm\theta_{21}\,,\quad
 \bm\zeta=\frac{\varepsilon_3\varepsilon_4}{\omega}\bm\theta_{34}\,               \nonumber\\
&\bm Q=\bm\Delta_\perp-\bm\Delta_0\,,\quad I_0=\frac{2 (\bm Q_\perp\cdot\bm\zeta)}{(M^2+\zeta^2)^2}\,,\quad \bm I_1= \frac{\bm Q_\perp}{M^2+\zeta^2}- I_0\bm\zeta\,.
\end{align}
Note that Eq.~\eqref{T1C} is valid for the region $Q\lesssim R^{-1} \ll m_\mu$ (where $R$ is the nuclear radius), which gives the main contribution to the cross section integrated over $\bm p_{3\perp}$ and $\bm p_{4\perp}$. In this region the Coulomb corrections to the amplitude $J^{(1)}$ of $\mu^+\mu^-$ pair production by virtual photon are absent. Besides, screening is important only for very high energies,
$$
\varepsilon_1\gtrsim \frac{m_\mu^2}{\alpha Z^{1/3} m_e}\sim 1\, \text{TeV} \,,
$$
and we  neglect this effect in our consideration. 

Then, for the sake of simplicity of calculations, we use the model potential 
\begin{equation}\label{V}
V(r)=-\frac{\eta}{\sqrt{r^2+R^2}}\,.
\end{equation}
For this potential, the form factor $F(Q^2)$ and the function $A(\bm \Delta_\perp)$ have a simple forms
\begin{align}
F(Q^2)=QR\, K_1(QR)\,,\quad A(\bm\Delta_\perp)=A_{as}(\bm\Delta_\perp) \frac{(\Delta_{\perp} R)^{1-i\eta}K_{1-i\eta}(\Delta_{\perp} R)}{2^{-i\eta}\Gamma(1-i\eta)}\,, 
\end{align}
where $K_\nu(x)$ is a modified Bessel function of the second kind and $A_{as}(\bm\Delta_\perp)$ is given in Eq.~\eqref{ACJCas}.

In fact, the difference between the results obtained by using the  realistic form factor and the model one is about $10\%$ (see Ref.~\cite{JS09} where the Born case has been considered). A small influence of the potential shape is irrelevant for the qualitative analysis of the importance of the Coulomb corrections related to the electron-atom interaction. 

\begin{figure}[h]
	\centering
	\includegraphics[width=0.49\linewidth]{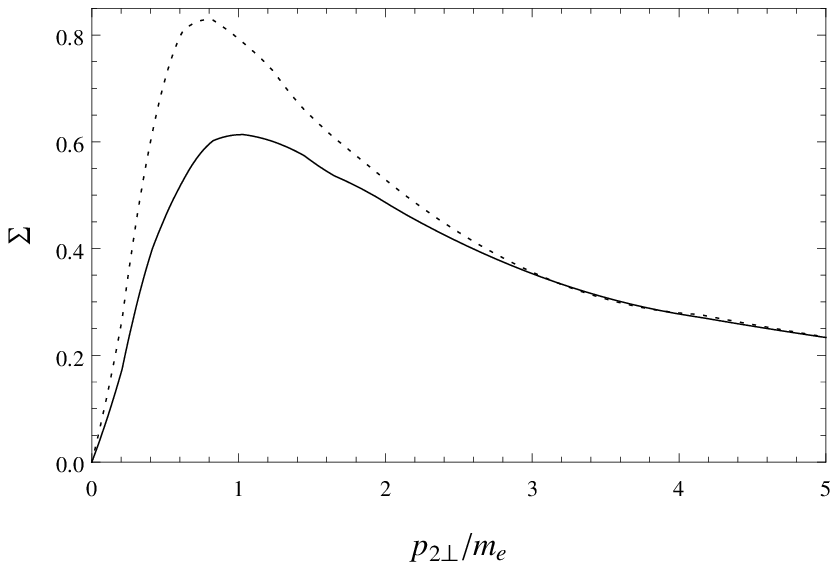}
	\includegraphics[width=0.49\linewidth]{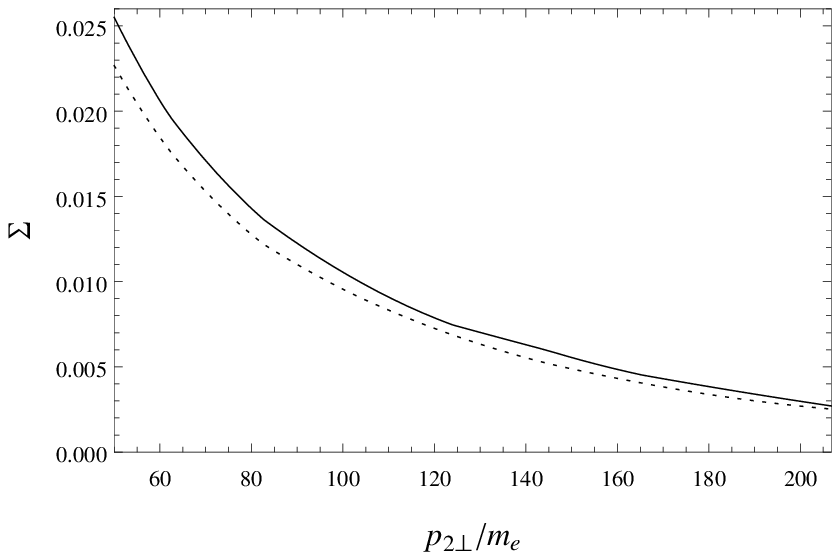}
	\caption{The dependence of $\Sigma$, Eq.~\eqref{Sig},  on $p_{2\perp}/m_e$ for $\omega=\varepsilon_1/2$, $\varepsilon_3=\varepsilon_4=\omega/2$,  $\varepsilon_1=50 m_\mu$, $Z=79$ (gold); solid curve is the exact result  and  dotted curve is the Born result.}
	\label{difpee}
\end{figure}

Let us consider the dimensionless quantity $\Sigma$,
\begin{equation}\label{Sig}
\Sigma=\frac{d\sigma}{Sdp_{2\perp}d\varepsilon_3d\varepsilon_4}\,, \quad S=\frac{\eta^2}{\omega^2m_\mu^2 m_e}\,,
\end{equation}
which is the differential cross section, integrated over $\bm p_{3\perp}$ and $\bm p_{4\perp}$,  in units $S$. This quantity is shown in Fig.~\ref{difpee} as the function of  $p_{2\perp}$ for
$\omega=\varepsilon_1/2$, $\varepsilon_3=\varepsilon_4=\omega/2$,  $\varepsilon_1=50\, m_\mu$, $Z=79$ (gold).

It is seen that the impact of the electron interaction with the atomic field on the cross section, differential over $p_{2\perp}$, is  significant.
In the region  $p_{2\perp}\sim m_e$,  the exact cross section is essentially smaller than that obtained in the Born approximation (the deviation of the Born result from the exact one is about $20-30\,\%$ ). At  $p_{2\perp}\gg m_e$, the exact cross section is lager than the Born one (the deviation  is about $10 \%$).  

\begin{figure}[H]
	\centering
	\includegraphics[width=0.65\linewidth]{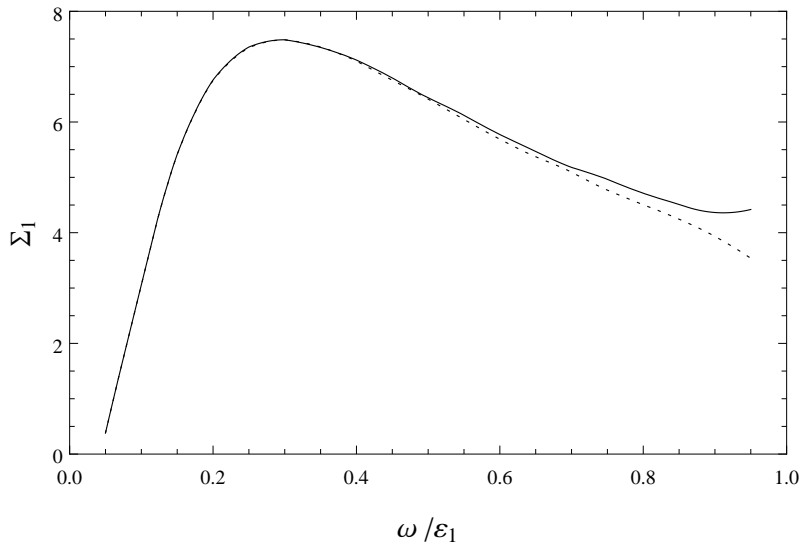}
	\caption{The dependence of $\Sigma_1$, Eq.~\eqref{sigma1},  on $\omega/\varepsilon_1$ for $\varepsilon_3=\varepsilon_4=\omega/2$,  $\varepsilon_1=50 m_\mu$, $Z=79$ (gold); solid curve is the exact result and dotted curve is the result obtained by neglecting the contribution of $T^{(0)}$.}
	\label{int_p2}
\end{figure}

In our resent paper \cite{KM2017},  the process of $e^+e^-$ pair production by a heavy particle in the atomic field has been discussed. In that paper it is shown that the cross section, differential with respect to the momentum of a heavy particle, is  strongly affected by the interaction of this particle with the atomic field. However, the cross section integrated over the final momentum of a heavy particle is  independent of this interaction.
It turns out that a similar statement is also true for the process of $\mu^+\mu^-$ pair production by a high-energy electron in the atomic field, i.e., the Coulomb corrections to the quantity
\begin{equation}\label{sigma1}
\Sigma_1=\frac{1}{m_e}\int_0^\infty \Sigma\, dp_{2\perp}
\end{equation}
are strongly suppressed. In Fig.~\ref{int_p2}, the solid curve shows  the quantity $\Sigma_1$ as the function of $\omega/\varepsilon_1$. Due to the strong suppression of the Coulomb corrections, the exact result coincides with the Born one for all $\omega$. It is interesting to consider the relative contribution of the amplitude $T^{(0)}$ to the cross section. Compared to the term $T^{(1)}$, the term $T^{(0)}$ contains the suppression factor $\omega/\varepsilon_1$ at $\omega\ll \varepsilon_1$.
This is why the contribution of $T^{(0)}$ to the cross section is important only for $\omega\sim \varepsilon_1$. This statement is confirmed by Fig.~\ref{int_p2} where  $\Sigma_1$, obtained  by neglecting the contribution of $T^{(0)}$, is shown as the dotted curve. 

 The account for the term $T^{(0)}$ results in the asymmetry of the differential cross section 
 with respect to the permutation of $\mu^+$ and $\mu^-$ momenta, $\bm p_4\leftrightarrow\bm p_3$. Due to  the relations
 \begin{align}
 &T^{(0)}_{\mu_1\mu_2\mu_3\mu_4}(\bm p_2,\bm p_3,\bm p_4)=T^{(0)}_{\mu_1\mu_2\mu_4\mu_3}(\bm p_2,\bm p_4,\bm p_3)\,,\nonumber\\
 &T^{(1)}_{\mu_1\mu_2\mu_3\mu_4}(\bm p_2,\bm p_3,\bm p_4)= -T^{(1)}_{\mu_1\mu_2\mu_4\mu_3}(\bm p_2,\bm p_4,\bm p_3)\,,
\end{align}
 this asymmetry arises as a result of interference between $T^{(0)}$ and $T^{(1)}$. 
 
 Let us consider the cross section integrated over $\bm p_{2\perp}$, $d\sigma(\bm p_{3},\bm p_{4})$, and define  the asymmetry ${\cal A}$ as
 \begin{equation}\label{Asym}
 {\cal A}=\frac{d\sigma(\bm p_{3},\bm p_{4})-d\sigma(\bm p_{4},\bm p_{3})}{d\sigma(\bm p_{3},\bm p_{4})+d\sigma(\bm p_{4},\bm p_{3})}\,.
 \end{equation}
 
\begin{figure}[H]
 	\centering
 	\includegraphics[width=0.7\linewidth]{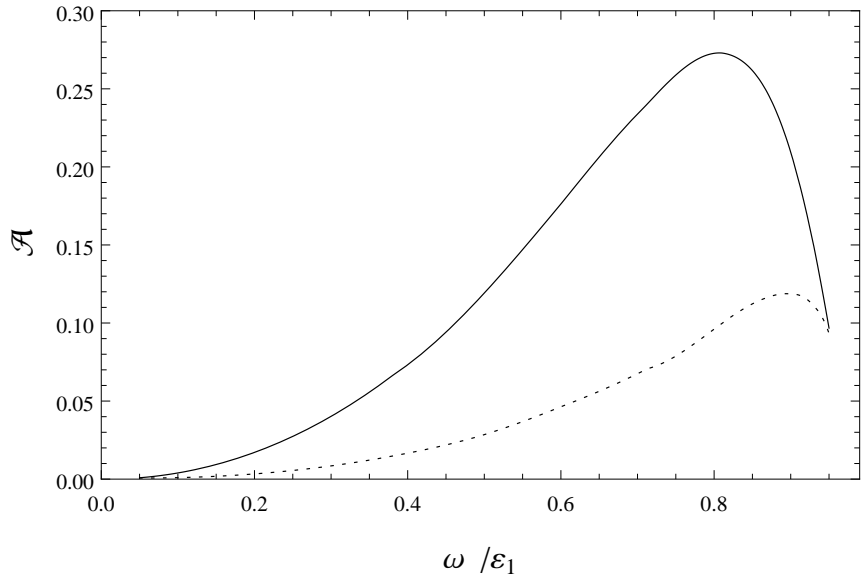}
 	\caption{The dependence of ${\cal A}$, Eq.~\eqref{Asym},  on $\omega/\varepsilon_{1}$ for,  $\varepsilon_1=50 m_\mu$, $\varepsilon_3=\varepsilon_4=\omega/2$, $\bm p_{3\perp}||-\bm p_{4\perp}$, $p_{4\perp}=m_\mu$,  $Z=79$ (gold); solid curve is the exact result for $p_{3\perp}=2.5 m_\mu$, dotted curve for $p_{3\perp}=0.5 m_\mu$.}
 	\label{As}
 \end{figure}

 In Fig.~\ref{As}, the asymmetry $\cal A$ is shown as the function of $\omega/\varepsilon_1$ for a few values of $\bm p_3$ and $\bm p_4$. It is seen that the asymmetry may reach tens of percent at $\omega\sim\varepsilon_1$.
 
 In Ref.~\cite{DLMR14}, the charge asymmetry of the cross section of  $\mu^+\mu^-$ photoproduction in the atomic field has been investigated. The asymmetry arises due to the account for the first quasiclassical correction to the amplitude of the process. The cross section of photoproduction calculated in the leading quasiclassical approximation does not possess such an asymmetry. The question arises whether it is possible to use  a beam of ultra-relativistic electrons as a source of equivalent photons for observation of the charge asymmetry in photoproduction, related to the next-to-leading quasiclassical approximation.  Since the charge asymmetry due to interference of the amplitudes $T^{(0)}$ and $T^{(1)}$ at  $\omega\sim\varepsilon_1$ may be large, the observation of the charge asymmetry in the process of electroproduction, which appears  due to the account for the next-to-leading quasiclassical corrections to the  amplitude of $\mu^+\mu^-$ pair production by a virtual photon,  becomes problematic.

\section{Paradimuonium electroproduction}\label{PDM}

In this section we consider the electroproduction of  $\mu^+\mu^-$ pair in the bound state by ultrarelativistic electron in the atomic field (dimuonium). In this process a dimuonium is mainly produced in the state with the total spin zero (paradimuonium with the positive C-parity) because in this case the amplitude $T^{(1)}$ is determined by  one virtual photon exchange of $\mu^+\mu^-$ pair with the atomic center. To produce an orthodimuonium (the total spin one and negative C-parity), it is necessary to have either two virtual photon exchange of a pair with the atomic center in the amplitude $T^{(1)}$, which is suppressed by the atomic form factor, or to account for only the amplitude $T^{(0)}$, which is small compared to the amplitude $T^{(1)}$.

The cross section $\sigma_{PM}$ of  high-energy paradimuonium electroproduction with the total angular momentum $l=0$ and the principal quantum number $n$ has the form~\cite{MSS71,Olsen85}
\begin{equation}\label{eq:cspara}
d\sigma_{PM}=\frac{\alpha^2\omega}{(2\pi)^5m_\mu}|\psi_n(0)|^2\,d\omega\,d\bm p_{2\perp}\,d\bm  P_\perp\,\frac{1}{2}\sum_{\mu_1\mu_2}|\widetilde{T}_{\mu_1\mu_2}|^{2}\,,
\end{equation}
where $\omega$ and $\bm P$  are the energy and the momentum of dimuonium, respectively, $\psi_n(0)$ is the dimuonium wave function at the origin, and $|\psi_n(0)|^2=\dfrac{\alpha^3 m_\mu^3}{8\pi n^3}$. The amplitude $\widetilde{T}_{\mu_1\mu_2}(\bm p_1,\bm p_2,\bm P)$ is expressed via the amplitude ${T}_{\mu_1\mu_2\mu_3\mu_4}^{(1)}(\bm p_1,\bm p_2,\bm p_3,\bm p_4)$, see Eq.~\eqref{T1C}, as follows
\begin{equation}
\widetilde{T}_{\mu_1\mu_2}(\bm p_1,\bm p_2,\bm P)={T}_{\mu_1\mu_2+-}^{(1)}(\bm p_1,\bm p_2,\bm P/2,\bm P/2)\,.
\end{equation}
We have
\begin{align}\label{T1CPM}
&\widetilde{T}=\frac{4i\eta\varepsilon_1}{\omega} \int\frac{d\bm\Delta_\perp\, A(\bm\Delta_\perp) F(Q^2)}{Q^2  M^2\,(m_e^2\omega^2+\varepsilon_1^2Y^2)}
{\cal M}\,, \nonumber\\
&{\cal M}=-\mu_1\delta_{\mu_1\mu_2}\big[\varepsilon_1 (\bm s_{\mu_1}^*\cdot\bm Y)(\bm s_{\mu_1}\cdot\bm Q_\perp)-\varepsilon_2(\bm s_{\mu_1}\cdot\bm Y)(\bm s_{\mu_1}^*\cdot\bm Q_\perp)  \big]+\delta_{\mu_1\bar\mu_2}\frac{m_e\omega^2}{\sqrt{2}\varepsilon_1 }(\bm s_{\mu_1}
\cdot\bm Q_\perp)\,,\nonumber\\
&M^2=m_\mu^2+\frac{\omega^2m_e^2}{4\varepsilon_1\varepsilon_2}
+\frac{\varepsilon_1}{4\varepsilon_2}Y^2\,,\quad \bm Y=\bm\Delta_\perp-\varepsilon_2\bm\theta_{21}\,,\quad
\bm Q=\bm\Delta_\perp+\bm p_1-\bm p_2-\bm P\,.
\end{align}

The cross section of paradimuonium pair production has similar properties as the cross section of unbound $\mu^+\mu^-$ pair production. I.e., the cross section differential over the electron transverse momentum $p_{2\perp}$ has the large Coulomb corrections, in contrast to the  commonly accepted point of view \cite{ACS00,GJKKSS98,KKSS99,HKS07}. To illustrate this statements we plot the dependence of the dimensionless quantity  $\Sigma_{PM}$,
\begin{equation}\label{SigPM}
\Sigma_{PM}=\frac{d\sigma_{PM}}{S_{PM}dp_{2\perp}d\omega}\,, \quad S_{PM}=\frac{\alpha^3\eta^2\zeta(3)}{8\pi\omega m_\mu^2 m_e}\,,
\end{equation}
on $p_{2\perp}$ for $Z=79$ and $\varepsilon_1=50 m_\mu$. In this formula, the summation over the principle quantum number is performed. 

\begin{figure}[h]
	\centering
	\includegraphics[width=0.49\linewidth]{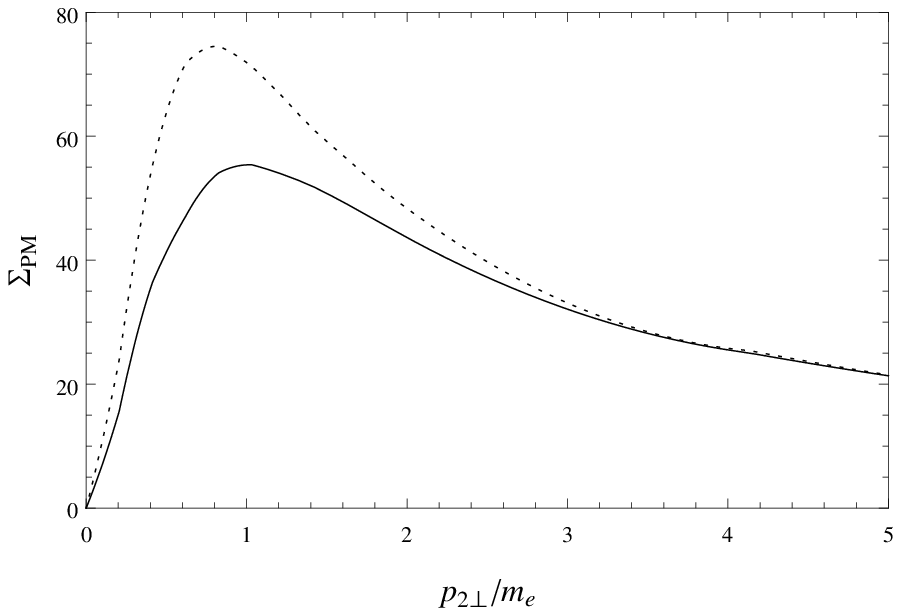}
	\includegraphics[width=0.49\linewidth]{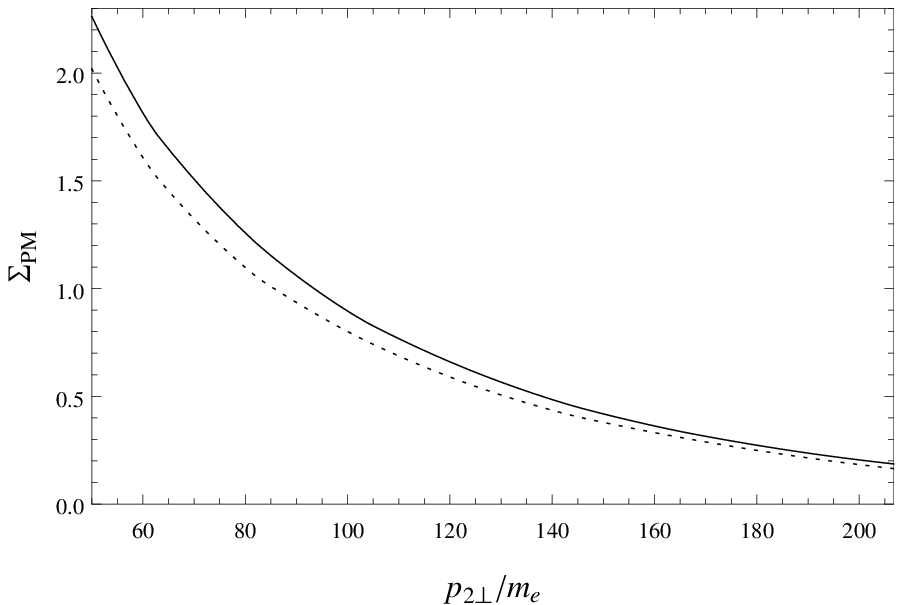}
	\caption{The dependence of $\Sigma_{PM}$, Eq.~\eqref{SigPM},  on $p_{2\perp}/m_e$ for $\omega=\varepsilon_1/2$, $\varepsilon_1=50 m_\mu$, $Z=79$ (gold); solid curve is the exact result  and  dotted curve is the Born result.}
	\label{difpeePM}
\end{figure}

\begin{figure}[H]
	\centering
	\includegraphics[width=0.7\linewidth]{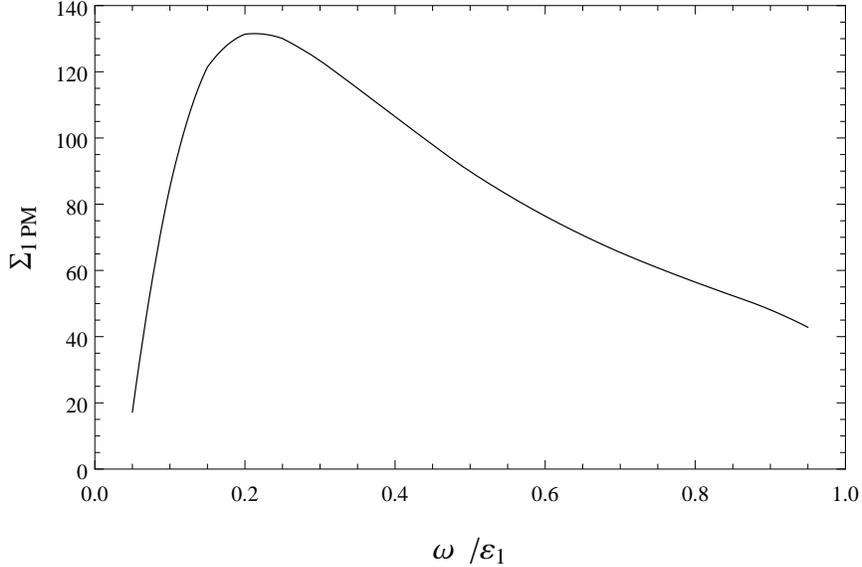}
	\caption{The dependence of $\Sigma_{1PM}$, Eq.~\eqref{Sig},  on $\omega/\varepsilon_{1}$ for,  $\varepsilon_1=50 m_\mu$, $Z=79$ (gold); solid curve is the exact result.}
	\label{int_p2PM}
\end{figure}

It is seen that the exact result in the peak region is about $30\%$ less than the Born result. In the wide region  $m_e\ll p_{2\perp}\lesssim m_\mu$ the exact result is about $10\%$ larger  than the Born one. Again, after integration over the momentum $ p_{2\perp}$, the cross section coincides with the Born result, see   Fig.~\ref{int_p2PM}, where the quantity 
$$\Sigma_{1PM}=\frac{1}{m_e}\int d p_{2\perp} \Sigma_{PM}$$
is shown as the function of $\omega/\varepsilon_1$ for $Z=79$ and $\varepsilon_1=50 m_{\mu}$.
The dependence of $\Sigma_{1PM}$ on $\omega/\varepsilon_1$ is very similar to that  shown in Fig.~\ref{int_p2} for $\Sigma_{1}$.
 
\section{Conclusion}\label{conc}
We have investigated the cross sections of $\mu^+\mu^-$ pair production by a high-energy electron in the strong atomic field. The interaction of electron with the atomic field is taken  into account exactly in the parameter $\eta$. The cases of the bound (paradimuonium) and unbound produced $\mu^+\mu^-$ pair are considered. For the cross sections differential over the transverse electron momenta $\bm p_{2\perp}$, the Coulomb corrections, related to the electron interaction with the atomic field, turns out to be large, in contrast to the commonly accepted point of view. Apparently, this effect can be easily observed experimentally. However, the Coulomb corrections to the cross sections integrated over $\bm p_{2\perp}$ are small. The asymmetry of the cross section with respect to the permutation  $\bm p_{3}\leftrightarrow \bm p_{4}$ of the momenta of $\mu^-$ and $\mu^+$ is  large. This asymmetry appears due to interference of the amplitudes $T^{(0)}$ and $T^{(1)}$ corresponding to production of a pair with the opposite  C-parity. This effect makes problematic a possibility to use an  electron beam as a source of equivalent photons for the charge asymmetry observation in $\mu^+\mu^-$ photoproduction in an atomic field. The latter asymmetry appears  due to account for the next-to leading quasiclassical corrections to the photoproduction amplitude.

\section*{Acknowledgement}
This work has been supported by Russian Science Foundation (Project No. 14-50-00080).

\section*{Appendix}
Here we present the Born amplitude $T_B$ for the process  of high-energy $\mu^+\mu^-$ electroproduction by an electron in the atomic field. In this case, the terms $T^{(0)}_{B\perp}$ and $T^{(0)}_{B\parallel}$ are given by Eq.~\eqref{T0} with the replacement 
$$
A(\bm\Delta_0)\rightarrow A_B(\bm\Delta_0)=-\frac{4\pi\eta}{\Delta_0^2}F(\Delta_0^2)\,.
$$
To derive the terms $T^{(1)}_{B\perp}$ and $T^{(1)}_{B\parallel}$, we use the following relation
\begin{equation}\label{BEq}
\lim_{\eta\rightarrow 0}\, A(\bm \Delta_\perp)=i(2\pi)^2 \, \delta(\bm \Delta_\perp)\,.
\end{equation}
Then we obtain
\begin{align}\label{T1CB}
&T_{B\perp}^{(1)}=-\frac{32\pi^2\eta\varepsilon_1\, F(\Delta_0^2)}{\omega\Delta_0^2 \,(m_e^2\omega^2+\varepsilon_1^2 p_{2\perp}^2)}{\cal M}_B\,, \nonumber\\
&{\cal M}_B=\frac{\delta_{\mu_1\mu_2}\delta_{\mu_3\bar\mu_4}}{\omega} \big[ \varepsilon_1(\varepsilon_3 \delta_{\mu_1\mu_3}-\varepsilon_4 \delta_{\mu_1\mu_4})
(\bm s_{\mu_1}^*\cdot\bm  p_{2\perp})(\bm s_{\mu_1}\cdot\bm I_{B1})\,\nonumber\\
&+\varepsilon_2(\varepsilon_3 \delta_{\mu_1\bar\mu_3}-\varepsilon_4 \delta_{\mu_1\bar\mu_4})(\bm s_{\mu_1}\cdot\bm  p_{2\perp})(\bm s_{\mu_1}^*\cdot\bm I_{B1})  \big]+\delta_{\mu_1\bar\mu_2}\delta_{\mu_3\bar\mu_4}\frac{m_e\omega\mu_1}{\sqrt{2}\varepsilon_1 }(\varepsilon_3 \delta_{\mu_1\mu_3}-\varepsilon_4 \delta_{\mu_1\mu_4})(\bm s_{\mu_1}
\cdot\bm I_{B1})\nonumber\\
&-\delta_{\mu_1\mu_2}\delta_{\mu_3\mu_4}\frac{m_\mu\mu_3}{\sqrt{2}}(\varepsilon_1 \delta_{\mu_1\mu_3}+\varepsilon_2 \delta_{\mu_1\bar\mu_3})(\bm s_{\mu_3}^*\cdot\bm  p_{2\perp})I_{B0} 
-\frac{m_em_\mu\omega^2}{2\varepsilon_1}\delta_{\mu_1\bar\mu_2}
\delta_{\mu_3\mu_4}\delta_{\mu_1\mu_3}I_{B0}\,,\nonumber\\
&T_{B\parallel}^{(1)}=\frac{32\pi^2\eta\varepsilon_3\varepsilon_4\,F(\Delta_0^2)}{\omega^3\Delta_0^2}\,I_{B0}
\delta_{\mu_1\mu_2}\delta_{\mu_3\bar\mu_4}\,, \nonumber\\       
&I_{B0}=-\frac{2(\bm\Delta_{0\perp}\cdot\bm\zeta)}{(M_B^2+\zeta^2)^2}\,,\quad \bm I_{B1}= -\frac{\bm\Delta_{0\perp}}{M_B^2+\zeta^2}- I_{B0}\bm\zeta\,,\quad \bm{\Delta}_0=\bm p_2+\bm p_3+\bm p_4-\bm p_1\,,\nonumber\\
&M_B^2=m_\mu^2+\frac{\varepsilon_3\varepsilon_4}{\varepsilon_1\varepsilon_2}m_e^2
+\frac{\varepsilon_1\varepsilon_3\varepsilon_4}{\varepsilon_2\omega^2} p_{2\perp}^2\,,\quad \bm  p_{2\perp}=\varepsilon_2\bm\theta_{21}\,,\quad
\bm\zeta=\frac{\varepsilon_3\varepsilon_4}{\omega}\bm\theta_{34}\,.  
\end{align}

The Born amplitude  $\widetilde{T}_B$ of the paradimuonium electroproduction (see Eq.\eqref{T1CPM})  can also be obtained by means of  Eq.~\eqref{BEq}:
\begin{align}\label{T1CPM}
&\widetilde{T}_B=-\frac{16\pi^2\eta\varepsilon_1F(\Delta_0^2)}{\omega\Delta_0^2  M^2_B\,(m_e^2\omega^2+\varepsilon_1^2p_{2\perp}^2)}
{\cal M}_B\,, \nonumber\\
&{\cal M}_B=\mu_1\delta_{\mu_1\mu_2}\big[\varepsilon_1 (\bm s_{\mu_1}^*\cdot\bm p_{2\perp})(\bm s_{\mu_1}\cdot\bm \Delta_{0\perp})-\varepsilon_2(\bm s_{\mu_1}\cdot\bm p_{2\perp})(\bm s_{\mu_1}^*\cdot\bm \Delta_{0\perp})  \big]+\delta_{\mu_1\bar\mu_2}\frac{m_e\omega^2}{\sqrt{2}\varepsilon_1 }(\bm s_{\mu_1}
\cdot\bm \Delta_{0\perp})\,,\nonumber\\
&M^2_B=m_\mu^2+\frac{\omega^2m_e^2}{4\varepsilon_1\varepsilon_2} +\frac{\varepsilon_1}{4\varepsilon_2}p_{2\perp}^2\,,\quad \bm \Delta_0= \bm p_2+\bm P-\bm p_1\,.
\end{align}

 \end{document}